\documentclass[reprint,aps, twocolumn, superscriptaddress]{revtex4}
\usepackage{graphicx}
\usepackage{dcolumn}
 \usepackage[mathlines]{lineno}
\usepackage{hyperref}
\usepackage{bm}
\usepackage{amssymb}
 \usepackage{braket}
 \usepackage{amsmath}
 \usepackage{color}
 \usepackage{soul} 
 \usepackage{graphics}

\begin{document}
\title{Vector mode decay in atmospheric turbulence: a quantum inspired analysis}
\author{Isaac Nape}
 \affiliation{School of Physics, University of the Witwatersrand, Private Bag 3, Johannesburg 2050, South Africa}
 \author{Nikiwe Mashaba}
 \affiliation{School of Physics, University of the Witwatersrand, Private Bag 3, Johannesburg 2050, South Africa}
 \affiliation{Optronic Sensor Systems, Defence and Security, Council for Scientific and Industrial Research (CSIR), Pretoria 0001, South Africa}
  \author{Nokwazi Mphuthi}
 \affiliation{School of Physics, University of the Witwatersrand, Private Bag 3, Johannesburg 2050, South Africa}
 \author{Sruthy Jayakumar}%
 \affiliation{Department of Electrical Engineering, Indian Institute of Technology Madras, Chennai – 600036, India}
\author{Shanti Bhattacharya}%
 \affiliation{Department of Electrical Engineering, Indian Institute of Technology Madras, Chennai – 600036, India}
\author{Andrew Forbes}%
 \email{andrew.forbes@wits.ac.za }
\affiliation{School of Physics, University of the Witwatersrand, Private Bag 3, Johannesburg 2050, South Africa}%

\begin{abstract}
\noindent Vector beams are inhomogeneously polarized optical fields with nonseparable, quantum-like correlations between their polarisation and spatial components, and hold tremendous promise for classical and quantum communication across various channels, e.g. the atmosphere, underwater, and in optical fibre. Here we show that by exploiting their quantum-like features by virtue of the non-separability of the field, the decay of both the polarisation and spatial components can be studied in tandem. In particular, we invoke the principle of channel state duality to show that the degree of nonseparability of any vector mode is purely determined by that of a maximally nonseparable one, which we confirm using orbital angular momentum (OAM) as an example for topological charges of $\ell=1$ and $\ell =10$ in a turbulent atmosphere.  A consequence is that the well-known cylindrical vector vortex beams are sufficient to predict the behaviour of all vector OAM states through the channel, and find that the rate of decay in vector quality decreases with increasing OAM value, even though the spread in OAM is opposite, increasing with OAM. Our approach offers a fast and easy probe of noisy channels, while at the same time revealing the power of quantum tools applied to classical light.
\end{abstract}

\date{\today}
\maketitle

\section{Introduction}

\noindent Structured light has become topical of late \cite{rubinsztein2016roadmap}, with so called cylindrical vector vortex (CVV) beams \cite{zhan2009cylindrical} taking centre stage in numerous fundamental and applied studies \cite{ndagano2017creation, rosales2018review}.  For example, they form a family of natural solutions of free-space and optical waveguides, and have been used in optical trapping  \cite{li2019optical, zhang2010trapping, bhebhe2018vector}, metrology \cite{berg2015classically}, as well as high capacity classical \cite{milione20154, zhao2015high,  willner2018vector} and quantum \cite{de2020experimental, sit2017high, cozzolino2019orbital, parigi2015storage} communication. To meet the demand of such growing applications, a plethora of generation methods have emerged, including directly from lasers \cite{sun2012low, naidoo2016controlled, sroor2020high}, or externally with liquid crystal q-plate technology \cite{marrucci2013q}, metasurfaces \cite{devlin2017arbitrary, mueller2017metasurface}, and spatial light modulators \cite{rosales2020polarisation}.  Detection has likewise matured to include deterministic detectors incorporating interferometers \cite{nagali2010experimental, slussarenko2010polarizing}, mode sorters \cite{walsh2016pancharatnam} or both \cite{ndagano2017deterministic}, as well as fast digital Stokes measurements \cite{selyem2019basis,manthalkar2020all,singh2020digital} and direct measures of the nonseparability or vector quality factor \cite{mclaren2015measuring,Ndagano2016}, giving a quantitative measure of how vector the vector beam is.

An open challenge in the context of classical and quantum communication is the propagation of such modes through media exhibiting spatially dependent perturbations.  These might include thermal effects due to overheating of optical elements \cite{whinnery1967thermal}, rapid refractive index fluctuations in the atmosphere \cite{andrews2005laser,cheng2009propagation} and underwater \cite{hufnagel2020investigation}, and in optical fibre \cite{ndagano2015fiber, chen2017theoretical}.  In particular, atmospheric turbulence leaves the polarisation of optical beams undisturbed while the spatial components degrade rapidly resulting in modal scattering and therefore information loss \cite{roux2011infinitesimal, ibrahim2013orbital, tyler2009influence, smith2006two, leonhard2018protecting, pors2011transport, malik2012influence, cox2016resilience}. Consequently, information encoding with the spatial components of light is restricted to only several km \cite{krenn2015twisted, krenn2014communication}. In the case, of vector beams, having coupled polarisation and spatial components, the polarisation fields are indirectly impacted \cite{chen2016characterizing}, resulting in the decay of nonseparability or vectorness \cite{ndagano2017characterizing}.

Here we exploit parallels between nonseparability of vector beams and entanglement in quantum systems \cite{ndagano2017characterizing,aiello2015quantum,toppel2014classical,forbes2019classically,Spreeuw1998,Eberly2016,konrad2019quantum,toninelli2019concepts} to deploy a quantum toolkit for the study of vector beams in atmospheric turbulence.  Importantly, we recognise that just as the degree of entanglement of any pure quantum state can be determined by that of a maximally entangled state \cite{konrad2008evolution}, courtesy of the Choi-–Jamiołkowski isomorphism (channel state duality) \cite{jamiolkowski1972linear, choi1975completely}, so it must be true that the dynamics of any vector beam in a one-sided noisy channel should be able to be inferred from the dynamics of just one beam, a purely inhomogeneously polarized field (a perfectly ``vector'' beam with orthogonal spatial modes), which from now on we will refer to as ideal vector vortex (VV) beams. In the context of OAM, any one of the CVV beams with oppoiste spin and OAM states would suffice, as well as VV beams in the linear polarisation basis.  Using such beams as a probe, we confirm the Choi-–Jamiołkowski isomorphism for classical vectorial light, and show that the vectorness decay of all initial beams can be predicted from the decay of an ideal VV beam. The approach is first outlined using CVV beams
and then generalised to other VV beams for adaptability. We illustrate this for two OAM subspaces, $\ell= \pm1$ and $\ell=\pm10$, revealing from this measure a simple factor for the rate at which one subspace decays relative to the other.  Our work not only offers a simple tool for probing classical communication channels, but also reveals insights into the decay dynamics of vectorial OAM light. While we have demonstrated the approach with OAM in the atmosphere, it can easily be adapted to other mode sets and media, and likewise to hybrid entangled quantum states.

\section{Concepts}

Here we elucidate the concept of channel state duality with nonseparable vector beams for characterising classical beams through perturbing media. We first introduce the key concepts of nonseparability in vector beams, vector beam decay through turbulence and then finally channel state duality.  The core idea is that an ideal vector beam is sufficient to predict the behaviour of any vector beam, even partially vector, through a channel.
\begin{figure}[h]
\centering
\includegraphics[width=1\linewidth]{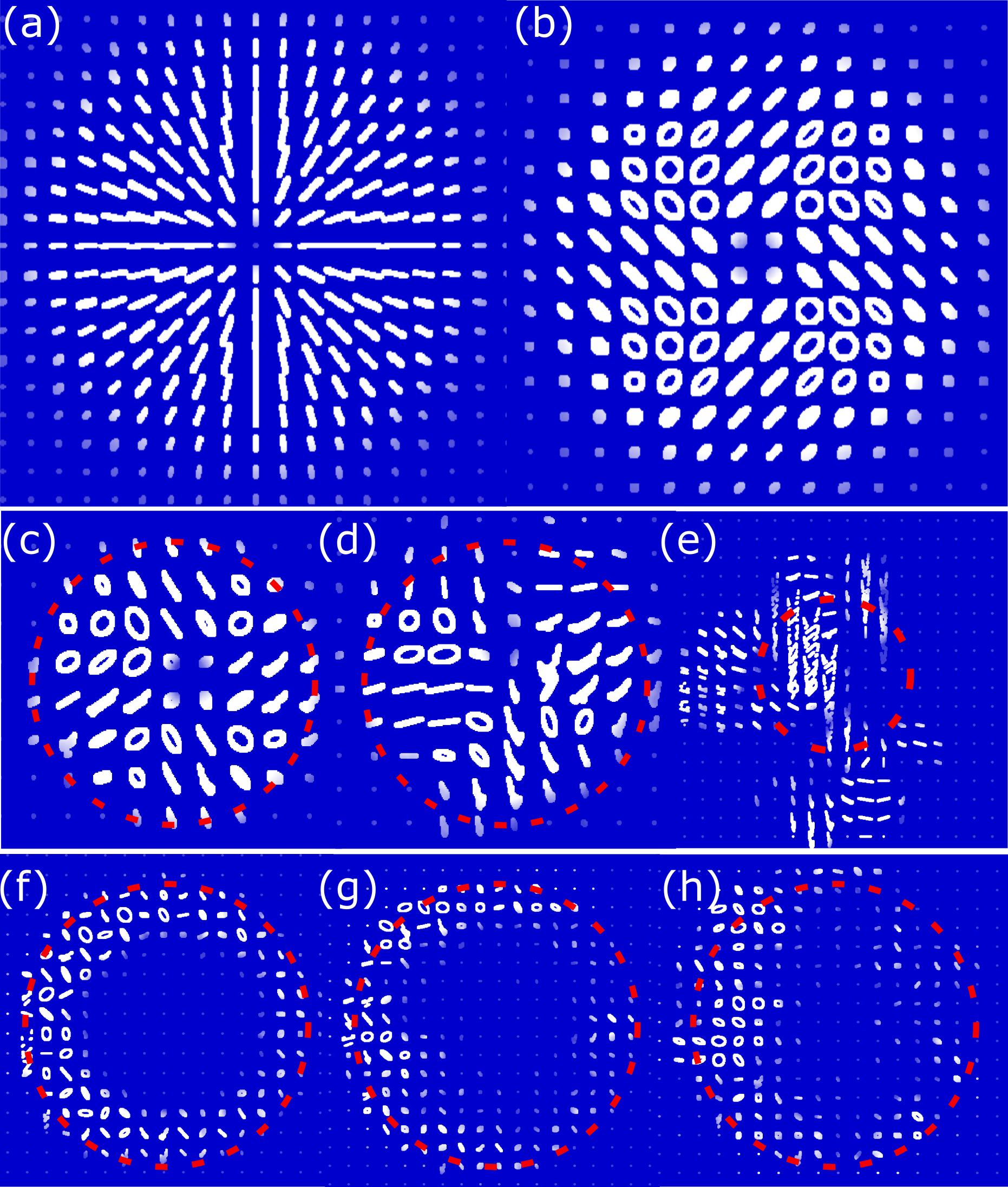}
\caption{ Example polarisation field profiles of a (a) CVV beam  (b) VV beam. The CVV beam only has nonuniform linear polarisation states while the VV beam varies between linear and elliptical polarisation states. Both beams have a VQF of 1. Experimental polarisation fields of vector modes transmitted through turbulence with varying strengths from left to right, initially encoded  in OAM  subspaces (c)-(e) $|\ell|=1$  and  (f)-(h) $|\ell|=10$ in the second and third bottom panels, respectively. The turbulence strengths are $D/r_0=0$ (no turbulence), $D/r_0=2.5$ and $D/r_0=3.5$ in each column, respectively.  }
\label{fig:Polmappings}
\end{figure}
\subsection{Nonseparable vector modes}
To construct an ideal VV beam we require a vectorial superposiiton of spatial mode and polarization where the nonseparability is maximum.  We illustrate the concept of nonseparability using familiar CVV beams since they are ubiquitous in a myriad of applications \cite{rosales2017simultaneous}. Thereafter, we extend the concept to a family of VV beams by a simple change of basis in the polarisation components.
 
 CVV beams are natural solutions of the vectorial Helmholtz equation in cylindrical coordinates. An example of one such beam, is shown in Fig.~\ref{fig:Polmappings}~(a), having a radially symmetric polarisation field. Such beams are commonly represented as superpositions of scalar fields coupled to orthogonal circular polarisation states, i.e.,

\begin{equation}
 \Psi_{\ell}(\textbf{r}) = a \exp(i \ell \phi)\hat{e}_{R} + b \ \exp(-i \alpha) \exp(-i \ell \phi)\hat{e}_{L},
  \label{eq : VectM}
\end{equation}
where $\textbf{r} = (r, \phi, z=0)$ are the cylindrical coordinates,  $\hat{e}_{R,L}$ are the canonical right and left circular polarisation states and coupled to spatial components having characteristic azimuthal phase profiles, $\exp( \pm i  \ell \phi)$, associated with light fields carrying an OAM of $\pm \ell \hbar$ per photon,  respectively. Here, the unbounded integer, $\pm\ell$, is the topological charge.  The parameters, $a$, $b$ and $\alpha$ are the relative amplitudes and phases between the the modes in the superposition. Next, we show that the coupling between the polarisation and spatial components is reminiscent of quantum entanglement between two particles.

\begin{figure*}[t]
\centering
\includegraphics[width=\linewidth]{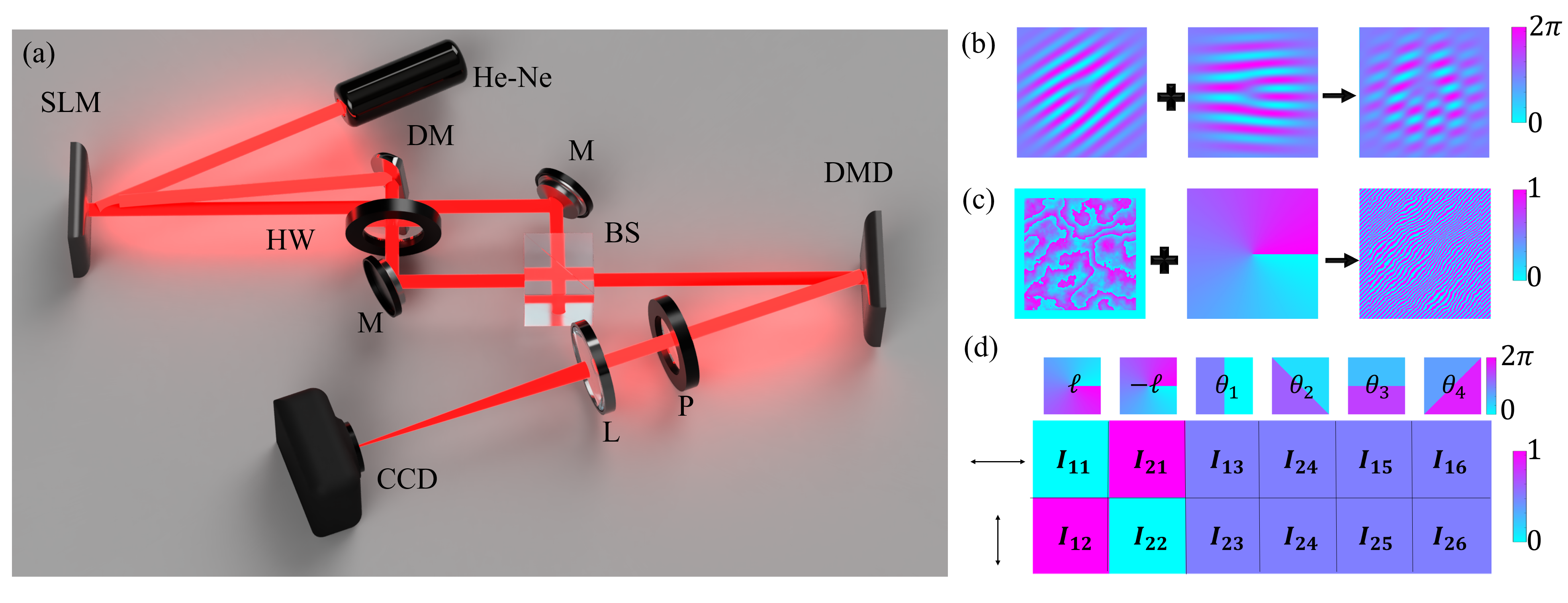}
\caption{(a) Illustration of the experimental set up. A Helium Neon (He-Ne) laser was expanded and collimated onto a spatial light modulater (SLM). On the SLM, two modes with oppositely charged OAM and distinct gratings frequencies were encoded on a single hologram as shown in (b). Upon propagation, the modes separated in path. Since they propagate closely, a D--shaped mirror (DM) as used to redirect one of them. A half-wave (HW) plate  was used to rotate the polarisation of redirected beam. The modes were subsequently recombined at a beam-splitter (BS). The resulting mode was imaged to the digital micromirror device (DMD) where (c) turbulence was encoded in combination with the detection holograms. Polarisation projections were performed with a linear polariser. Finally, the resulting mode was propagated to the far field with a 500 mm Fourier lens (L) where an on-axis intensity measurement was performed with a CCD camera. (d) Example of measurements needed to calculate the nonseparability, VQF, of vector modes.}
\label{fig:setup}
\end{figure*}
It is common to represent the polarisation and spatial components of vector modes as state vectors using the Dirac notation from quantum mechanics, i.e., $\hat{e}_{R (L)} \rightarrow \ket{R (L)}$ and $\exp(\pm i\ell \phi) \rightarrow \ket{\pm \ell}$, enabling a more compact representation of Eq.~(\ref{eq : VectM}) following
\begin{align}
\ket{\Psi_{\ell}}= a \ket{\ell} \ket{R} +  b \exp(-i \alpha) \ket{-\ell} \ket{L}.
\label{eq : VectM2}
\end{align}
Here the bra-ket notation is used to mark the spatial and polarisation components making it convenient to express each Degree of Freedom (DoF) as a unique subsystem analogous to two particle states in quantum mechanics. By clearly identifying each DoF, we wish to quantify the amount of nonseparability between them. This can be achieved by using the vector quality factor (VQF)~\cite{mclaren2015measuring,Ndagano2016}, which is an analogous measure of entanglement based on the concurrence \cite{wootters2001entanglement}, but between the internal DoF of the classical light fields. For the state in Eq.~(\ref{eq : VectM2}), the VQF, equivalently concurrence, is therefore given by $\text{VQF}=|ab|$ ranging from $\text{VQF} = 0 $ for a separable scalar beam $(a=0 \ \text{or} \ b=0)$ to  $ \text{VQF} = 1 $ for a nonseparable vector beam ($a=b$); and otherwise partially vector for $0<\text{VQF} <1$. This measure has been used in a myriad of experiments as a witness for nonseparability in classical beams \cite{sroor2018purity, otte2018recovery, selyem2019basis}. 

Note that such beams are spanned on a four dimensional state-space and therefore have the general form
\begin{align}
\ket{\Phi_{\ell}} &=  a\ket{\ell} \ket{R} + b \ket{-\ell} \ket{R} \nonumber \\
&+c \ket{\ell} \ket{L}  + d\ket{-\ell} \ket{L} \big),
\end{align}
with a corresponding degree of nonseparability
\begin{equation}
    \text{VQF} = |ad - cb|,
\end{equation}
assuming the coefficients satisfy, $(|a|^2  +|b|^2 + |c|^2 +  |d|^2) = 1$. One can simply recover the previous vector beam for example by setting ($a=b$) while $cb = 0$. This will become crucial when studying the decay of vector beams since we will always project onto the entire subspace ($\pm \ell$) of spatial modes we started with.

Next, we show that the coupling between the DoFs of any spatial mode that is transmitted through a complex medium, using turbulence as an example, can be determined by that of a maximally nonseparable vector mode by exploiting channel state duality.\\ 

\subsection{Propagation of vector modes through turbulence and channel state duality}

\begin{figure*}[t]
\centering
\includegraphics[width=\linewidth]{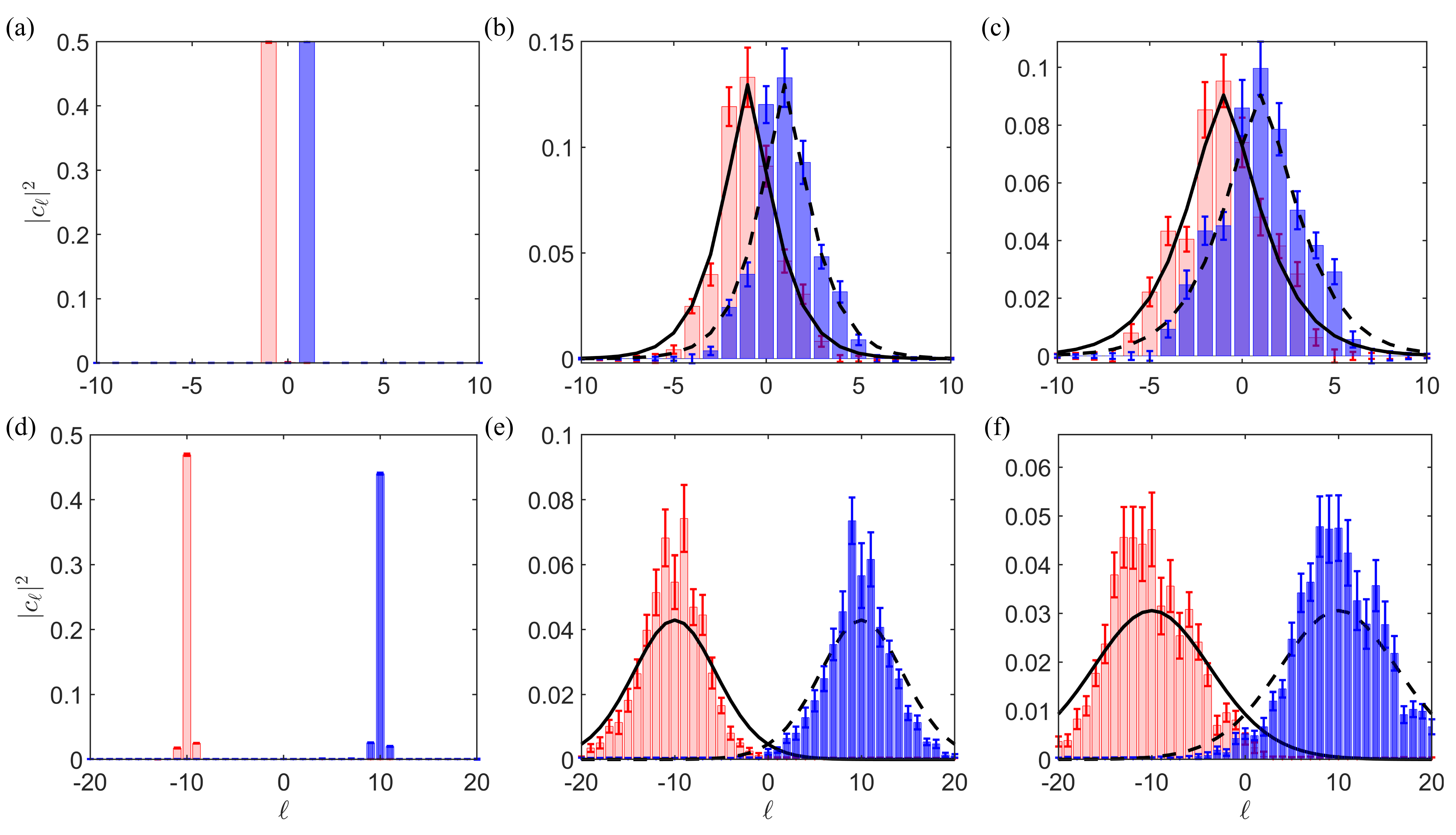}
\caption{Measured modal spectrum for the vertical and horizontal polarised components of vector modes in OAM subspaces corresponding to $\ell=1$ (top panel) and $\ell=10$ (bottom panel), with (a,d) no  turbulence $D/r_0 = 0$  and increased turbulence strengths of (b,e) $D/r_0 = 2.5$ and  (c,f) $D/r_0 = 3.5$. The solid lines correspond to the theoretical spectrum.}
\label{fig:spectrum}
\end{figure*}

Structured light is known to be perturbed in atmospheric turbulence \cite{cox2020structured}.  In particular, a vector mode propagating through turbulence experiences phase dependent fluctuations that have an impact on its transverse spatial components. Assuming weak irradiance fluctuations approximated by Kolmogorov theory \cite{andrews2005laser}, the phase variations can be characterised by the phase structure function, $D_{\phi}(\mathbf{r}_1,  \mathbf{r}_2)= 6.88\Delta r / r_0$, where $r_0$ is the Fried parameter \cite{fried1965statistics} describing the transverse scale of the atmospheric distortions and $\Delta r = |\mathbf{r}_1-\mathbf{r}_2|$ are relative displacements in the transverse plane. For an optical system with a diameter (aperture) $D$, we can associate the turbulence strength with the normalised aperture size, $D/r_0$, relating turbulence strength to the relative transverse distance within which the refractive index is correlated. For example a large aperture size \textit{seeing} a smaller Fried parameter ($D>r_0$) experiences more distortions than a smaller aperture \textit{seeing} a larger Fried parameter ($D<r_0$). In this paper, $D$ approximates the size of the beam. We show examples of the the effect of turbulence on the polarisation field of vector modes in Fig.~\ref{fig:Polmappings} (c)-(h). The polarisation field as well as the spatial distribution gets deformed with increasing turbulence strength (from left to right). We investigate how the distortions affect the nonsepability.

Now, since the atmosphere is non-birefringent, only the transverse spatial components of the mode expressed in Eq.~(\ref{eq : VectM2}) are perturbed and as a consequence there is modal scattering into adjacent OAM modes \cite{chen2016characterizing}. For example, an OAM mode corresponding to the state $\ket{\ell}$ traversing a medium with the channel matrix $\hat{T} = \sum_{m, n} c_{n, m} \ket{m}\bra{n}$, transfers energy from the initial state into its neighbouring eigenmodes following the mapping 
\begin{equation}
\ket{\ell} \xrightarrow{\hat{T}} \sum_{m}c_{\ell, m}\ket{m}, 
\label{eq : modescattering}
\end{equation}
where $P(\ell , m) = |c_{\ell, m}|^2$ is the conditional probability for the mode $\ket{\ell}$ to exchange energy with the mode $\ket{m}$. For Kolmogorov turbulence, the  probabilities have been determined analytically \cite{zhang2020mode} and are symmetric about $|\ell|$ for weak turbulence. 

With this in mind, we can now describe the state of our CVV mode after it traverses the turbulent channel. Let us assume that vector mode is initially in the state
\begin{align}
\ket{\Psi_{\ell}} =  1/\sqrt{2} \left( \ket{\ell} \ket{R} +  \ket{-\ell} \ket{L} \right).
\label{eq : VectM3}
\end{align}
Upon traversing the channel, the OAM modes scatter according to Eq.~(\ref{eq : modescattering}). By projecting back onto the initial OAM subspace, i.e., $\{ \ket{-\ell}, \ket{\ell} \}$, we obtain the state  

\begin{align}
\ket{\tilde{\Psi}_{\ell}} &= \mathcal{N} \big( c_{\ell, \ell}\ket{\ell} \ket{R} + c_{\ell, -\ell}\ket{-\ell} \ket{R} \nonumber \\
&+ c_{-\ell, \ell} \ket{\ell} \ket{L}  + c_{-\ell, -\ell}\ket{-\ell} \ket{L} \big).
\label{eq : VectMTurb}
\end{align}

\noindent Here $\mathcal{N}$ is a normalisation factor satisfying, $|\braket{\tilde{\Psi}_{\ell}|\tilde{\Psi}_{\ell}}|^2  = 1$. The VQF becomes

\begin{equation}
\mbox{VQF}_{\text{max}} = 2|\mathcal{N}|^2|c_{\ell, -\ell}c_{-\ell, \ell}-c_{\ell, \ell}c_{-\ell, -\ell}|.
\end{equation}

By applying the notion of the Choi-–Jamiołkowski isomorphism (channel state duality), we hypothesis that any other vector mode with initial vectorness of $\mbox{VQF}_{\text{in}}$ will decay according to the factorisation law, 

\begin{equation}
\mbox{VQF}_{\text{out}} = \mbox{VQF}_{\text{max}} \times \mbox{VQF}_{\text{in}}
\label{maineq}
\end{equation}

\noindent where $\mbox{VQF}_{\text{out}}$ is the nonseparability of the state after the channel. In other words, the decay of the any arbitrary vector beam can be inferred by simply propagating an ideal vector beam through the channel to find the scale factor, $\mbox{VQF}_{\text{max}}$. Rather than propagating many beams through the channel, only one beam has to be passed through to understand its impact.  Remarkably, the channel's properties are determined by its interaction with a maximally nonseparable vector mode. This means that by knowing how a maximally nonseparable vector mode propagates through turbulence, it is possible to predict how the nonseparability of any other arbitrary superposition state evolves. Equation (\ref{maineq}) also predicts that the trend is linear and an intercept at zero, with the decay of the CVV beam returning the slope ($\mbox{VQF}_{\text{max}}$).

This intriguing property, can be understood from the perspective of quantum mechanics. Firstly, the channel weights, $|\braket{\pm\ell,|\hat{T}|\ell}|^2$, are imprinted on the input nonseparable state mapping the channel onto a pure state, as a consequence of the Choi--Jamiołkowski isomorphism \cite{choi1975completely, jamiolkowski1972linear}. Equivalently, a vector mode traversing a noisy channel has similar properties owing to the nonseparability of the spatial and polarisation components. This means that the final VQF of any partial nonseparable mode can be determined by that of a maximally nonseparable vector mode.

Finally, although we have used the CVV beam as a well-known example, in general the polarisation and spatial components can be in any basis.  For convenience going forward and in the experiment, we will convert to the horizontal ($\ket{H}$) and vertical ($\ket{V}$) polarisation basis while the spatial profiles are defined in the Laguerre-Gaussian (LG) basis. As a result, we no longer have a CVV beam since there is no cylindrical symmetry in the polarisation field. An example of our VV beam is shown in Fig.~\ref{fig:Polmappings} (b), where the polarisation states across the tranverse plane are mixtures of linear of elliptical states. Such a beam is represented by the state,
\begin{align}
\ket{\Psi_{\ell}} &=  a \ket{\ell} \ket{H} + b \ket{-\ell} \ket{V},
\label{eq:VVbeam}
\end{align}
with the same VQF = 1 as that of Eq.~(\ref{eq : VectM2}). This serves to make it clear than any ideal VV is sufficient for the test.

\section{Methods}
\subsection{Experimental set-up}
We describe the generation and detection scheme illustrated in Fig.~\ref{fig:setup} (a). We used a Helium-Neon (He-Ne) laser with a central wavelength of 633 nm and collimated Gaussian field profile. We modulated the laser beam using a Phase only Holo-Eye Pluto spatial light modulate (SLM). To obtain the states $\ket{\Psi}_{\ell=1, 10}$, we encoded the multiplexed holograms \cite{rosales2017simultaneous} with Laguerre Gaussian (LG) modes of charges $\ell=-1 (-10)$ and $\ell=1(10)$ and subsequently separated them in path using a D shaped mirror. The amplitudes and phases of each mode were encoded using the Arrizon technique for complete amplitude and phase control \cite{arrizon2007pixelated}. An example of one of the holograms is shown in Fig.~\ref{fig:setup} (b).

Before interfering the two beams at the BS, we rotated the polarisation of the reflected beam from the D--shape mirror by $45^\circ$ using a half wave-plate. This converted the polarisation from H to V. The two beams now had orthogonal polarisations. After combining the two beams, the resulting vector mode was transmitted to the digital micro-mirror device (DMD). On the DMD, we encoded Kolmogorov turbulence phase screens following \cite{zhao2012aberration} in combination with the detection holograms necessary for the VQF measurements. An example of one of the detection holograms is shown in Fig.~\ref{fig:setup} (c) as a combination of the detection mode and turbulence phase screen, resulting in a noisy detection hologram that has both the perturbation from turbulence and projection mode. The VQF projection holograms, with no turbulence, had phase profiles shown in the first row of Fig.~\ref{fig:setup} (d), shown for the $\ell=\pm1$ subspace. 

Lastly we used a polariser to project onto the H and V polarisation modes after the DMD. The resulting field was then propagated to the far field using a Fourier lens (L) and an on axis measurement of the intensity was recorded, providing the modal overlap of the input state, the simulated turbulence and the detection mode \cite{flamm2012mode}. For each measurement we prepared up to 30 instances of each turbulence strength ranging from $D/r_o$=$0$ to $3.5$ in steps of $D/r_o=0.5$.
An example of measurements for intensities $I_{uv}$ is shown in Fig.~\ref{fig:setup} (d) for a perfect vector mode. The columns correspond to the spatial projections while the rows correspond to the polarisation measurements. Next we show how the VQF (nonseparability) is measured.

\subsection{Vector quality factor measurement}

We follow the procedure outlined in ref.~\cite{mclaren2015measuring} to measure the nonseparability of vector modes. The VQF is given by
\begin{equation}
\text{VQF} = \sqrt{1-\sum^{3}_{i}\braket{\sigma_{i}}^2}
\label{eq : VQF},
\end{equation}
\noindent where the expectation values of the Pauli matrices $\langle\sigma_i \rangle$ can be obtained from
\begin{align}
&\langle\sigma_1\rangle = I_{13}+I_{23}-(I_{15}+I_{25}),\\
&\langle\sigma_2\rangle = I_{14}+I_{24}-(I_{16}+I_{26}),\\
&\langle\sigma_3\rangle = I_{11}+I_{21}-(I_{12}+I_{22}).
\end{align}
The detection probabilities  $I_{uv}$, $u = \lbrace1,2\rbrace$ , $v = \lbrace1,2,...,6\rbrace$ are determined from six identical projections of different polarisation basis states, namely horizontal and vertical polarisations. The projections are performed by inserting a polariser, set to $ 0^{\circ}$ and  $90^{\circ}$ for the horizontal and vertical polarisation projections, respectively. The six spatial measurements consist of projections onto OAM states $\ket{\pm\ell}$ and their four superpositions $\ket{\theta}=\ket{-\ell}+e^{i\frac{\theta}{|\ell|}}\ket{\ell}$ with $\theta = 0, \frac{\pi}{2}, \pi$ and $\frac{3\pi}{2}$. The spatial projections were encoded as binary holograms onto the digital micro-mirror device using the method in \cite{lee1979binary} tailored for amplitude only devices. Finally, the on-axis intensity, due to each projection is measured in the focal plane of a Fourier lens by a CCD camera, in the first diffraction order.

\section{Results}
\begin{figure*}[t]
\centering
\includegraphics[width=1\linewidth]{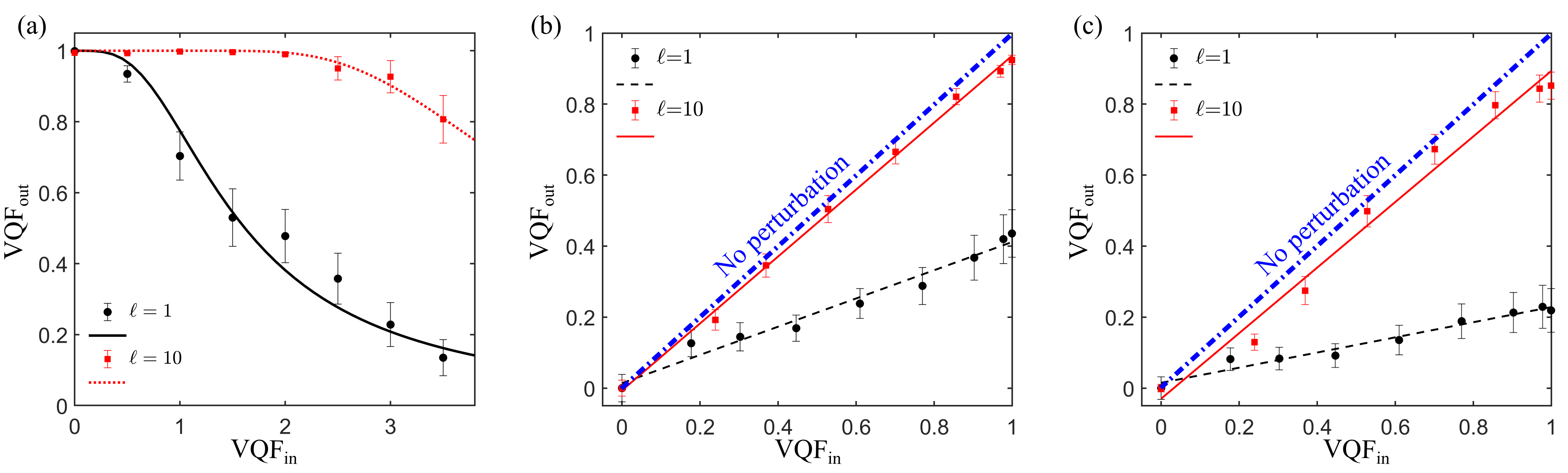}
\caption{(a) Experimental results (points) and theoretical prediction (lines) for the degree of nonsepability of vector modes in the subspaces, $\ell=1$ and $10$, with increasing turbulence strength. Experimental output VQF with respect to the known input VQF under the effect of turbulence strengths of (b) $D/r_0 =2.5$ and (c) $D/r_0 =3.5$. In the absence of perturbations, the output VQF maps onto the diagonal (``No perturbation'') line. The circles are for vector modes in the subspace of $\ell=1$ and squares are for $\ell=10$, while the lines are the theoretical prediction based on the isomorphism. The horizontal error-bars are smaller than the points. Each data point was obtained from 30 realisations of the same turbulence strength.}
\label{fig:choi}
\end{figure*}
 \subsection{Propagation of vector modes through turbulence}
We first generated vector modes ($\ket{\Psi_{\ell}}$) and measured their polarisations using Stokes polirometry. The polarisations profiles are shown in Fig.~\ref{fig:Polmappings} (c)-(e) and (f)-(h) for $\ket{\Psi_{1}}$ and $\ket{\Psi_{10}}$, respectively. In each panel, the profiles are shown for increasing turbulence strengths  $D = 0, 2.5$ and $D=3.5$, from left to right. The intensity profile is shown to deteriorate in each instance confirming the presence of distortions in the transverse plane of the fields. 

Next, we measured the modal spectrum of each polarisation component averaged over 30 instances of the same turbulence strength. The results are shown in Fig~\ref{fig:spectrum}~(a)-(c) for $\ket{\Psi_{\ell=1}}$  and in Fig~\ref{fig:spectrum}~(d)-(f) for  $\ket{\Psi_{\ell=10}}$. For each $\ell$ dependent mode, the turbulence strength was $D/r_0$ = 0 (a , d)  $D/r_0$=2.5 (b , e) and $D/r_0$=3.5 (c, f). In each plot, the distribution on the right corresponds the horizontally polarised mode (blue) while the distribution on the left (red) corresponds to the vertically polarised mode. As expected, the mode distribution is symmetric about $|\ell|$, consistent with the theoretical distribution shown as lines \cite{zhai2020turbulence}. We measured the width of each distribution as 2 times the standard deviation using the formula
\begin{equation}
   \Delta \ell = 2 \sqrt{ \frac{\sum_{m} |m-\overline{\ell}| P(\ell, m)}{\sum_{m} P(\ell, m)}},
\end{equation}
\noindent where $\overline{\ell}$ is the mean OAM in the field.
For $D/r_0=2.5$ we measured a width of $\Delta \ell = 3.18$ and  $\Delta \ell = 6.64$ for $\ket{\Psi_{\ell=1}}$  and $\ket{\Psi_{10}}$, respectively, averaged over both polarisation components of the beams. With an increased turbulence strength, $D/r_0= 3.5$, we measured a width of $\Delta \ell = 4.44$ and $\Delta\ell = 8.79$. In both cases the OAM width of $\ell=10$ vector modes is higher than that of $\ell=1$ vector modes (2 times), showing that higher order OAM modes spread farther than lower order OAM modes, consistent with theory \cite{chen2016characterizing}.

While this reveals information about how the spatial components are perturbed, we now investigate how the polarisation in tandem with the spatial components are affected, testing our isomorphism hypothesis. 

To illustrate the effect of the mode scattering on the nonseparability, we show the theoretical (lines) and experimental (points) of VQF values measured for  $|\ell|=1, 10$ subspaces as a function of turbulence strength in Fig.~\ref{fig:choi} (a). The VQF of the $\ell=10$ modes decay at lower rate than the VQF of the $\ell=1$ modes, analogous to the decay of entangled photons through a single sided channel of turbulence \cite{ibrahim2013orbital}. This can be explained by the larger mode separation between higher OAM suspaces enabling for less crosstalk between highly separated~\cite{malik2012influence}.
\begin{table}[h!]
  \begin{center}
    \caption{Analysis of the different OAM subspace through turbulence. The values in brackets correspond to the theoretical values. Here the gradients, $m_\ell$, correspond to $\mbox{VQF}_{\text{max}}$.}
    \label{tab:table1}
    \begin{tabular}{|c|c|c|c|c|c|c|c|}
    \hline
      $D/r_0$ & $\Delta\ell \ (\ell = 1)$ & $\Delta \ell \ (\ell=10)$ &  $m_{1}$ & $R^2_{1}$ & $m_{10}$ & $R^2_{10}$\\
      \hline
      0 &  0.05 (0)   & 0.84 (0)      & - & - & -& -\\
      2.5& 3.18 (4.1) & 6.64 (9)      & 0.40 & 0.97 & 0.94 & 0.99\\ 
      3.5& 4.44 (5.59) & 8.79 (11.94) & 0.21 & 0.97 & 0.92 & 0.98\\ 
      \hline
    \end{tabular}
    \label{table: Sammry_results}
  \end{center}
\end{table}
While the above analysis was performed on input modes with a high nonseparability ($VQF\approx1$), next we evaluated how vector modes with a varying degree of nonseparability also decay under the same turbulence strength. We demonstrate this to confirm the channel state duality inherent in vector modes propagated through a perturbation channel acting on the spatial DoF. The results are shown in Fig.~\ref{fig:choi} (b) and (c) under turbulence conditions of  $D/r_o = 2.5$ and $3.5$, respectively. This was done for subspaces $\ell=\pm1$ (circles) and $\ell=\pm10$ (squares).

The error-bars for the x-axis are smaller than the points. To control the VQF of the input mode, we adjusted the grating depth of the hologram corresponding to the $\ell=-1 (-10)$ mode. The plots show that the output VQF, i.e., $\mbox{VQF}_{\text{out}}$, has a linear relation to the input VQF where the line fitted through each data set has a gradient, $m_{\ell}$ or equivalently $\mbox{VQF}_{\text{max}}$, that is equivalent to the VQF of a maximally nonseparable vector mode transmitted through the same turbulence. We show the gradient of each line in Table \ref{table: Sammry_results} as well as the goodness of fit, which are above $R^2 = 0.96$. A perfect fit would result in $R^2=1$. As shown, the gradients of the $\ell=10$ modes is higher than that of $\ell=1$, owing to the higher crosstalk in the $\ell=1$ subspace as demonstrated earlier. For $D/r_o=3.5$, we find that the $\ell=10$ subspace, the gradient can be 4 times larger in comparison to the $\ell=1$ subspace. Since the gradient indicates the maximum VQF that a vector mode with an input nonseparability of $\mbox{VQF}_{\text{in}}\approx1$ can obtain, after propagating through the channel, all partially nonseparable vector modes within the same subspace are bounded on the interval $\mbox{VQF}_{\text{out}}\in[0,m_{\ell}]$ consistent with the factorisation law \cite{konrad2008evolution} for  single sided channels indicating that vector modes posses the ability to probe channel state duality of noisy channels. Indeed, our results show that the decay in nonseparability of any vector mode, in the same subspace decays through the medium according to the relation $\mbox{VQF}_{\text{out}} = \mbox{VQF}_{\text{in}}\times m_{\ell} =\mbox{VQF}_{\text{in}} \times \mbox{VQF}_{\text{max}}$ where $\mbox{VQF}_{\text{max}} = \mbox{VQF}(\hat{T} \ket{\Psi}_{\ell})$  is the VQF of a vector mode, $\ket{\Psi}_{\ell}$, after traversing the channel, $\hat{T}$, and having an initial VQF of 1 while $\mbox{VQF}_{\text{in}}$ is the VQF of the mode we wish to characterise after the channel. The isomorphism holds for vectorial classical light and can be used to probe classical channels.

\section{Discussion }
 
Vector beams possess nonseparable coupling between their polarisation and spatial components and exhibiting correlations similar to entangled pairs of photons. In this paper, we used this fact to study the decay of vector beams in atmospheric turbulence both qualitatively and quantitatively by invoking properties such as channel state duality \cite{choi1975completely, jamiolkowski1972linear} and the factorisation law \cite{konrad2008evolution}. These features are unique to quantum entangled states, and are commonly used for channel (medium) characterisation \cite{malik2010quantum}. Our results confirm that the nonseparability of any other partially ($\text{VQF} < 1$) nonseparable vector mode is purely determined by that of a maximally ($\text{VQF} \approx 1$) nonseparable vector mode experiencing the same turbulence within the same subspace of optical modes. Interestingly, this does not only limit the method to characterisation purposes but creates the possibility of using vector modes as a means to overcoming turbulence, for example by selecting higher order modes (e.g $\ell=10$ as apposed to $\ell=1$), or for unscrambling complex aberrations shown to be feasible for high dimensional quantum states \cite{valencia2020unscrambling}. The extension of the latter to classical beams could be used as an additional tool for adaptive optics since vector modes can carry information related to the channel/medium. While we demonstrated this method for turbulence, it can in principle be extended to various scenarios where optical aberrations are encountered, for example,  arising from overheating optical elements in high power regimes or imperfections in optical elements. 

\section{Conclusion}
In summary, we exploited the concept of channel state duality analogously to characterise the evolution of various vector/scalar modes through turbulence, demonstrating that the evolution of vector beams can be used to study how various spatial modes decay through turbulence. We showed this for two subspaces, $\ell=1$ and $\ell =10$, with our results demonstrating that higher order OAM vector modes decay rapidly while also maintaining a high nonseparability. Our work is integral to the development of alternative methods for characterising optical beams by borrowing principles from quantum mechanics, with possible applications in various scenarios where complex perturbations are encountered.

\section*{Appendix}
 \subsection{Scattering probability of OAM in turbulence}
 The detection probability of an OAM beam propagated through turbulence can computed from 
 \begin{equation}
     P(l,m) = \iint C_{\psi} (r, \Delta \theta, z) r dr \times \frac{\exp(-i m \Delta \theta)}{2 \pi} d \Delta \theta,
     \label{eq:TurbOverlap}
 \end{equation}
 \noindent where $\ell$ is the input OAM index of the beam, and $m$ is the index for the scattered mode, while  $C_{\psi}  \left( r, \Delta \theta, z \right)$ is the rotational coherence function defined in the cylindrical coordinates, $(r, \theta , z)$ defined as  \cite{paterson2005atmospheric}
  \begin{equation}
      C_{\psi}(r, \Delta \theta, z) = \braket{\psi^* (r, 0, z) \psi(r, \Delta \theta, z)}.
 \end{equation}
 Here $\psi(r, \theta, z) = u(r, \theta, z) \exp(i \phi(r, \theta))$ is the beam profile after propagating a distance $z$ with an initial profile of $u(r, \theta, 0)$ at the waist plane and $\exp(i\phi(r, \theta))$ is the accumulated phase according to the Rytov approximation. 
 
 For LG beams, the integral in Eq.~(\ref{eq:TurbOverlap}) has been solved analytically, yielding the in the expression \cite{zhai2020turbulence}
  \begin{align}
     P(l,m) &= (\frac{1}{t})^{|\ell|+1} \left(\frac{t-1}{t+1}\right)^{n}\sum_{k=0}^{|\ell|}  \binom{|\ell| + n }{k} \nonumber \\
            &\times \binom{2|\ell| - k }{\ell}\binom{4t }{(t-1)^2}^{k-|\ell|}
     \label{eq:TurbOverlapSoln}
 \end{align}
 \noindent where $n=|\ell-m|$, $t=\sqrt{1+\zeta}$ while $\zeta = 3.44 \times 2^{2/3}(w_0/r_0)^2$ where $w_0$ is the is the waist size of the Gaussian argument in the LG mode and $r_0$ is the Fried parameter. We re-scaled the Gaussian argument, i.e. $ w_\ell = w_0/\sqrt{|\ell|+1}$, so that each OAM mode has the same diameter, $D:=\sqrt{8} w_\ell$ \cite{andrews2005laser} and turbulence strength, $D/r_0$.
 

\end{document}